\documentclass[runningheads]{llncs}
\usepackage[T1]{fontenc}
\usepackage{graphicx}
\usepackage{inconsolata}
\usepackage{amsmath}
\usepackage{tikz}
\usetikzlibrary{shapes.geometric, arrows.meta, positioning, backgrounds, fit}
\usepackage[table]{xcolor}
\usepackage{booktabs}
\usepackage{colortbl}
\usepackage{subcaption}
\usepackage{listings}
\begin{document}
\title{Reflect-SQL: A Self-Reflection Based Framework for Text-to-SQL}
\titlerunning{Reflect-SQL: A Self-Reflection Based Framework}
%
\author{Anupreksha Jain\inst{1}\orcidID{0009-0004-5609-2108} \and
Manish Shrivastava\inst{1}\orcidID{0000-0001-8705-6637}}
\authorrunning{F. Author et al.}
%
\institute{International Institute of Information Technology Hyderabad, India \\
\email{anupreksha.j@research.iiit.ac.in}\\
\email{m.shrivastava@iiit.ac.in}}
%
%
\maketitle              
\begin{abstract}
Democratizing data access through natural language is a crucial goal for modern enterprises, but the practical adoption of Text-to-SQL is critically hindered by real-world complexities: 1. Obscure and large database schemas, 2. Ineffective retrieval of relevant tables and columns due to structured setting of schemas and vague user query, 3.  Generation of syntactically or logically flawed SQL due to a lack of robust validation and correction mechanism. To address these systemic challenges, we introduce Reflect-SQL, a novel framework for Text to SQL, grounded in multi-stage self-reflection approach to develop understanding of obscure schema using a knowledge base, setup a process for effective retrieval and system to generate syntactically/semantically SQL. Instead of a single-pass attempt, our system employs an LLM-as-a-judge driven scoring mechanism within interconnected feedback loops to iteratively refine the results at every stage. A feedback-driven retrieval loop refines the user's natural language query, while a synthesis loop validates and corrects the SQL and finally, an entailment loop optimizes the end-to-end process and continuously enriches the knowledge base. By integrating these layers of reflection, Reflect-SQL bridges the critical gap between user intent and complex data. On the challenging BIRD benchmark, our framework achieves an execution accuracy of 72.03\%, significantly outperforming state-of-the-art baselines, demonstrating a major leap in reliability for enterprise applications.

\end{abstract}

\section{Introduction}
Successfully deploying Text-to-SQL in enterprise environments hinges on overcoming three critical challenges:

1. Schema Obscurity: Deciphering large-scale databases characterized by cryptic column names (e.g., 'A2', 'dname') and undocumented relationships.

2. Contextual Retrieval: Accurately extracting relevant schema elements is difficult, as vague user queries often fail to map clearly onto complex, structured database environments.

3. Logical Integrity: Moving beyond syntax to ensure semantic accuracy, preventing the generation of logically flawed SQL that yields incorrect results.

We introduce \textbf{Reflect-SQL}, a framework that overcomes these challenges through multi-stage self-reflection. The system utilizes a dynamic \textbf{Knowledge Base} to disambiguate cryptic database structures, enabling a \textbf{Feedback-Driven Retrieval} process that refines vague user queries for precise context. Additionally, an \textbf{Iterative SQL Synthesis} loop validates and corrects the generated code, using a final entailment check to guarantee that the results match the user's original intent.

Our experiments validate this multi-layered, reflective approach, demonstrating a significant improvement in the accuracy of 11.05\% on complex benchmarks. The primary contributions of this work are:
\begin{enumerate}
    \item \textbf{A complete Text-to-SQL Framework:} A comprehensive, end-to-end system for enterprise Text-to-SQL tasks.

    \item \textbf{Dynamic Knowledge Base:} A learning mechanism that enriches schema understanding over time.

    \item \textbf{Reference-Free SQL Evaluation:} A novel methodology for validating SQL without relying on gold-standard labels.

    \item \textbf{Feedback-Driven Query Refinement:} An iterative process that aligns ambiguous user queries with the database schema.
\end{enumerate}
\section{Related work}

\cite{ji2023towards} introduced the concept of self-reflection in language models to progressively improve the factuality and consistency of generated responses, primarily for open-domain tasks. In their approach, the model scores its own output and iteratively refines it on the basis of identified errors. Our work adapts and extends this paradigm to the structured domain of Text-to-SQL generation—a task with fundamentally different constraints and evaluation criteria for Knowledge Base Construction, Information Retrieval and SQL Generation.

\subsection{Knowledge Base Construction}

Recent work has established that rich knowledge bases, built by aligning metadata with business glossaries \cite{lobo2023matching}, enhancing column descriptions \cite{gao2025automatic}, \cite{qin2024relational}, \cite{wretblad2024synthetic}, or injecting domain knowledge \cite{ma2024enhancing}, significantly improve Text-to-SQL performance. However, these methods primarily treat the knowledge base as a static asset created once. Our approach introduces an iterative refinement loop, where the knowledge base is continuously enriched based on successful query executions, allowing it to adapt to ambiguous, real-world applications.

\subsection{Information Retrieval}

Retrieval-Augmented Generation (RAG), introduced by \cite{lewis2020retrieval} has evolved to include iterative query refinement to improve context quality \cite{asai2023self}, \cite{chan2024rq}, \cite{zhou2025openrag}. We adapt this principle to the structured domain of databases by creating a schema aware, hierarchical retrieval process. Unlike standard RAG, our framework retrieves information iteratively from Knowledge base (tables, related tables, and columns) and uses a concrete feedback score to reformulate the user's natural language query if the retrieved context is insufficient or irrelevant.

\subsection{Text-to-SQL}
The Text-to-SQL field has evolved from early encoder-decoder architectures like Seq2SQL \cite{zhong2017seq2sql} and SQLNet \cite{xu2017sqlnet} to the current paradigm dominated by Large Language Models (LLMs). Modern in-context learning methods, such as DIN-SQL \cite{pourreza2023din} and SQLPrompt \cite{sun2023sqlprompt}, have significantly improved performance leveraging few-shot exemplars and chain-of-thought reasoning. Our framework not only builds on these foundations but also introduces critical layers of self-correction. First, we employ an iterative SQL synthesis loop that goes beyond syntax to correct logical errors based on a multi-dimensional semantic score. Second, as a final safeguard, we add an entailment check that validates if the query's results actually answer the user's intent. 

\section{Knowledge base construction and refinement}
\begin{figure}[ht]
\centering
\resizebox{\textwidth}{!}{%
\begin{tikzpicture}[node distance=1cm and 1.5cm, >=Stealth]
    \colorlet{processFill}{teal!15}
    \colorlet{processBorder}{black!60}
    \colorlet{decisionFill}{orange!20}
    \colorlet{decisionBorder}{black!60}
    
    \tikzstyle{block} = [rectangle, rounded corners, draw=processBorder, thick, text centered, minimum height=1.5cm, text width=3cm, fill=processFill]
    \tikzstyle{decision} = [diamond, draw=decisionBorder, thick, aspect=1.5, text centered, text width=2.5cm, fill=decisionFill]
    \tikzstyle{arrow} = [->, thick, black!70]
    \tikzstyle{container} = [rectangle, rounded corners, draw, dashed, inner sep=0.5cm, black!80]

    \node[block] (preprocessing) {\small Generate table descriptions based on table schema};
    \node[block, right=of preprocessing] (retrieval) {\small Retrieve relevant tables and columns};
    \node[block, right=of retrieval] (generation) {\small Generate and refine SQL Query};
    \node[decision, right=of generation] (entailment) {Entailed?};
    \node[block, right=of entailment] (postprocessing) {\small Improve Knowledge base with user query and generated SQL};

    \draw [arrow] (preprocessing) -- (retrieval);
    \draw [arrow] (retrieval) -- (generation);
    \draw [arrow] (generation) -- (entailment);
    \draw [arrow] (entailment) -- node[above] {Yes} (postprocessing);
    
    \draw [arrow] (entailment.north) -- ++(0, 1.2cm) -| node[above, pos=0.25] {No} (retrieval.north);
    
    \node[container, fit=(preprocessing), label=above:KB Construction] {};
    \node[container, fit=(retrieval), label=above:Retrieval Loop] {};
    \node[container, fit=(generation), label=above:SQL Generation Loop] {};
    \node[container, fit=(postprocessing), label=above:KB Enrichment] {};
\end{tikzpicture}
}
\caption{High-level architecture of the Reflect-SQL framework.}
\label{fig:high-level-arch-tikz}
\end{figure}
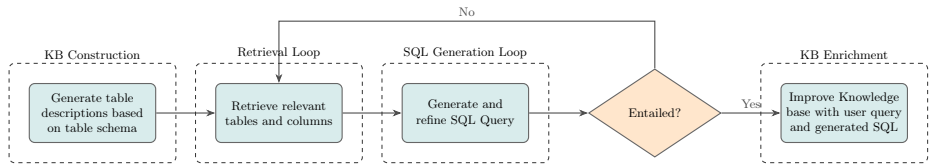

The knowledge base (KB) acts as a semantic abstraction layer containing table descriptions, column metadata, and relationships. This component is vital for handling industry-scale databases, which often exceed the context limits of Large Language Models (LLMs) and suffer from obscure naming conventions. By externalizing schema details into a structured KB, we overcome context constraints and limited schema understanding, ensuring the LLM accesses only the specific information required to bridge the gap between user intent and the database. The overall architecture of this framework is depicted in Fig.~\ref{fig:high-level-arch-tikz}.
\subsection{Construction}
The knowledge base serves as the semantic backbone of our framework. For the BIRD\cite{li2023bird} dataset, we construct this repository by synthesizing the database schema, sample data values, and the provided training set of sample questions and external evidence. In broader enterprise settings, this process also ingests domain-specific knowledge from business documents and glossaries. These inputs are processed to generate comprehensive descriptions for tables, columns, and their relationships (a detailed JSON representation of this structure is provided in Appendix~\ref{sec:appendix-kb-example}). To enable efficient retrieval, these components are encoded into dense vector representations using a sentence embedding model \cite{reimers2019sentence}. The vector database is structured hierarchically: a global collection for table descriptions and separate local collections for each table's column descriptions. This architecture supports a two-step retrieval process, first identifying relevant tables and subsequently narrowing down to specific columns.

\subsection{Iterative refinement}
In this post-processing step, after a successful SQL execution, if the query passes syntactic and semantic validations and entailment score between the query and result exceeds a threshold $\theta_e$, the knowledge base is refined. This conditional update is critical for preventing the propagation of errors. low-quality or contextually incorrect queries might pollute the knowledge base. Only syntactically and semantically validated queries with validated results contribute to the system's learning process.

This iterative update mechanism is essential in real-world environments, particularly those characterized by ambiguous or indecipherable column names and continuously evolving schemas. Example in Table~\ref{tab:kb-refinement-example} shows how the description of columns are refined after a successful query execution.

\begin{table}[ht]
\centering
\footnotesize 
\begin{tabular}{lp{4.5cm}p{6cm}}
\toprule
\textbf{Component} & \textbf{Before Refinement} & \textbf{After Refinement} \\
\midrule
\texttt{loan.status}    &  Categorical column for loan status with values 'A', 'B', 'C', 'D'.  & Loan repayment state: 'A' indicates finished or paid, 'B' indicates defaulted...  \\
\texttt{trans.k\_symbol} & Text column named 'k\_symbol' with values 'POJISTNE', 'SIPO'. & Transaction Category: determines the payment type (e.g. 'POJISTNE' maps to Insurance)... \\
\bottomrule
\end{tabular}
\caption{Example of Knowledge Base Refinement for BIRD\cite{li2023bird}}\label{tab:kb-refinement-example}
\end{table}
\section{Iterative RAG retrieval}
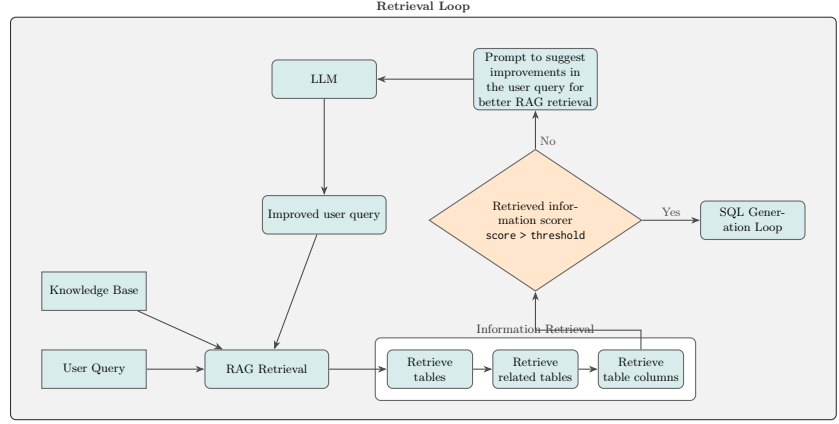
\begin{figure}[ht]
\centering
\resizebox{0.9\columnwidth}{!}{%
\begin{tikzpicture}[node distance=1cm and 0.5cm, >=Stealth]
    \colorlet{processFill}{teal!15}
    \colorlet{processBorder}{black!60}
    \colorlet{decisionFill}{orange!20}
    \colorlet{decisionBorder}{black!60}
    \colorlet{loopBg}{black!5}

    \tikzstyle{io} = [rectangle, draw=processBorder, thick, text centered, minimum height=1cm, text width=2.5cm, fill=processFill]
    \tikzstyle{proc} = [rectangle, rounded corners, draw=processBorder, thick, text centered, minimum height=1cm, text width=3cm, fill=processFill]
    \tikzstyle{decision} = [diamond, draw=decisionBorder, thick, aspect=1.5, text centered, text width=3.5cm, fill=decisionFill]
    \tikzstyle{arrow} = [->, thick, black!70]
    \tikzstyle{group} = [rectangle, rounded corners, draw=processBorder, thick, inner sep=0.3cm, label={[font=\small, black!80]above:Information Retrieval}]

    \node[io] (user_query) {User Query};
    \node[io, above=of user_query] (kb) {Knowledge Base};
    \node[proc, right=1.5cm of user_query] (rag) {RAG Retrieval};
    
    \node[proc, right=1.5cm of rag, text width=2cm] (tables) {Retrieve tables};
    \node[proc, right=of tables, text width=2cm] (related) {Retrieve related tables};
    \node[proc, right=of related, text width=2cm] (columns) {Retrieve table columns};
    
    \node[decision, above=1.5cm of related] (scorer) {Retrieved information scorer \\ \texttt{score > threshold}};
    
    \node[proc, above=1cm of scorer] (prompt) {Prompt to suggest improvements in the user query for better RAG retrieval};
    \node[proc, left=2.5 cm of prompt, text width=2.5cm] (llm) {LLM};
    \node[proc, below=2.5 cm of llm] (improved_query) {Improved user query};

    \node[proc, right=1.5cm of scorer, text width=2.5cm] (sql_gen_loop) {SQL Generation Loop};

    \draw [arrow] (user_query) -- (rag);
    \draw [arrow] (kb) -- (rag);
    \draw [arrow] (rag) -- (tables);
    \draw [arrow] (tables) -- (related);
    \draw [arrow] (related) -- (columns);
    
    \draw [arrow] (columns.north) -- ++(0,0.5cm) -| (scorer.south);

    \draw [arrow] (scorer.north) -- node[right, pos=0.2] {No} (prompt);
    \draw [arrow] (scorer.east) -- node[above] {Yes} (sql_gen_loop);
    
    \draw [arrow] (prompt) -- (llm);
    \draw [arrow] (llm) -- (improved_query);
    \draw [arrow] (improved_query) -- (rag);
    
    \begin{pgfonlayer}{background}
        \node[rectangle, rounded corners, draw=black!80, thick, fill=loopBg, fit=(user_query) (kb) (scorer) (prompt) (columns) (sql_gen_loop), inner sep=0.8cm, label={[font=\bfseries, black!80]above:Retrieval Loop}] {};
        \node[group, fill=white, fit=(tables) (related) (columns)] {};
    \end{pgfonlayer}
\end{tikzpicture}
}
\caption{Iterative Retrieval Loop.}
\label{fig:retrieval-loop-tikz}
\end{figure}

In the initial retrieval phase, we use the curated KB to identify relevant tables and columns for a user query, as detailed in Fig.~\ref{fig:retrieval-loop-tikz}.
\subsection{Table and column retrieval (hierarchical retrieval process)}
Given a user query $Q$, we embed it and use cosine similarity(as shown in \ref{eq:score_formula}) to retrieve the top-k tables $\{T_1, \dots, T_k\}$ from the vector database.
\begin{equation} \label{eq:score_formula}
\text{score}(T_i, Q) = \cos(\vec{t}_i, \vec{q}) = \frac{\vec{t}_i \cdot \vec{q}}{\|\vec{t}_i\| \|\vec{q}\|}
\end{equation}
This set is expanded by including tables related via relationships defined in the KB, ensuring join paths are available. For each selected table, a similar embedding-based search retrieves the most relevant columns.

\subsection{Relevance and completeness scoring}
We compute two metrics to assess retrieval quality: a \textbf{Relevance Score} ($R$) for semantic alignment and a \textbf{Completeness Score} ($C$) to ensure all necessary schema elements are present. These scores are generated by a novel approach of LLM-as-a-judge scoring (prompt details are added in appendix A). Final retrieval score ($Score_r$) is calculated with formula \ref{eq:retrieval_score}.
\begin{equation} \label{eq:retrieval_score}
Score_r = \alpha R + (1-\alpha) C , \quad \text{where } \quad \alpha \in [0,1]
\end{equation}

\subsection{Query Refinement}
If the retrieved schema is deemed insufficient ($Score_r < \theta_r$), an iterative refinement loop is triggered. In each iteration, the LLM first reformulates the user query into a more detailed version, $Q'$, to improve its semantic focus. Simultaneously, the retrieval scope is expanded by increasing the number of tables and columns to be retrieved in the next pass. This process of refining the query's focus while broadening the search area allows the system to quickly zero in on the complete set of schema elements required for complex queries. The loop terminates once the $Score_r$ exceeds the threshold $\theta_r$ or a maximum number of iterations is reached.

\section{SQL query generation}
\begin{figure}[ht]
\centering
\resizebox{\columnwidth}{!}{%
\begin{tikzpicture}[node distance=1.2cm and 1cm, >=Stealth]
    \colorlet{processFill}{teal!15}
    \colorlet{processBorder}{black!60}
    \colorlet{decisionFill}{orange!20}
    \colorlet{decisionBorder}{black!60}
    \colorlet{loopBg}{black!5}

    \tikzstyle{proc} = [rectangle, rounded corners, draw=processBorder, thick, text centered, minimum height=1cm, text width=4cm, fill=processFill]
    \tikzstyle{io} = [rectangle, draw=processBorder, thick, text centered, minimum height=1cm, text width=2.5cm, fill=processFill]
    \tikzstyle{decision} = [diamond, draw=decisionBorder, thick, aspect=1.5, text centered, text width=3cm, fill=decisionFill]
    \tikzstyle{arrow} = [->, thick, black!70]

    \node[proc] (prompt_initial) {Prompt based on given details (retrieved tables, columns, and relationships) to create SQL for the user query};
    \node[proc, right=1.5cm of prompt_initial, text width=2cm] (llm) {LLM};
    \node[io, below=of llm, text width=2cm] (sql_query) {SQL Query};
    \node[decision, below=of sql_query] (syntax_check) {SQL Syntactically correct?};
    \node[decision, right=1.5cm of syntax_check] (semantic_scorer) {SQL Semantic Scorer \\ \texttt{score > threshold}};
    \node[proc, above=3.2 cm of semantic_scorer, text width=5cm] (prompt_refine) {Prompt LLM to regenerate the accurate query based on semantic scorer feedback};
    \node[proc, left=1 cm of sql_query, text width=5cm] (prompt_refine_error) {Prompt LLM to regenerate the accurate query to correct the error in execution};
    
    \node[decision, right=1.5cm of semantic_scorer, text width=2.5cm] (entailment_check) {Entailment check};

    \draw [arrow] (prompt_initial) -- (llm);
    \draw [arrow] (llm) -- (sql_query);
    \draw [arrow] (sql_query) -- (syntax_check);
    \draw [arrow] (syntax_check) -- node[above] {Yes} (semantic_scorer);
    \draw [arrow] (prompt_refine_error.east) -- (llm);
    \draw [arrow] (syntax_check.west) -- ++(-2.4, 0) -- node[right, pos=0.8] {No}  (prompt_refine_error.south);
    \draw [arrow] (semantic_scorer.north) -- node[right, pos=0.2] {No} (prompt_refine);
    \draw [arrow] (prompt_refine) -- (llm);

    \draw [arrow] (semantic_scorer.east) -- node[above] {Yes} (entailment_check);
    
    \begin{pgfonlayer}{background}
        \node[rectangle, rounded corners, draw=black!80, thick, fill=loopBg, fit=(prompt_initial) (prompt_refine) (syntax_check) (semantic_scorer) (entailment_check), inner sep=0.8cm, label={[font=\bfseries, black!80]above:SQL Generation Loop}] {};
    \end{pgfonlayer}
\end{tikzpicture}
}
\caption{Iterative SQL Generation Loop.}
\label{fig:sql-gen-loop-tikz}
\end{figure}
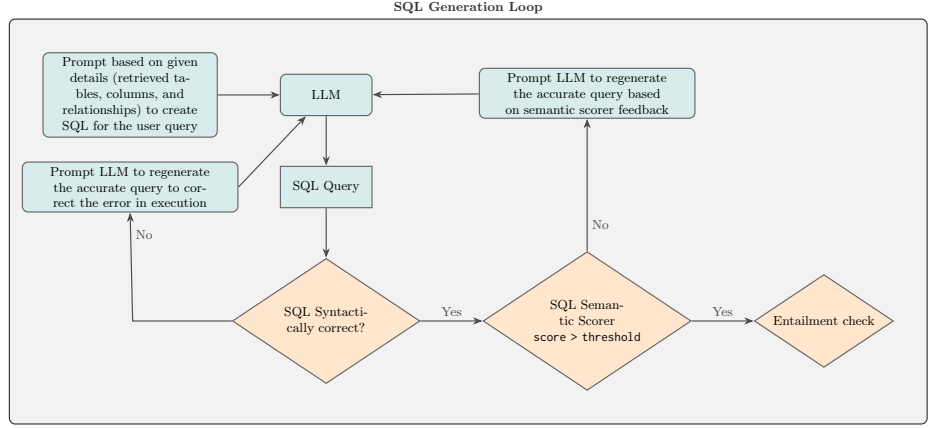

With a high-quality set of retrieved tables, columns, descriptions and sample data we generate a candidate SQL query using an LLM, as shown in Fig.~\ref{fig:sql-gen-loop-tikz}.
\subsection{Syntactic Validation and Correction Loop}
Each generated SQL query is immediately subjected to a syntactic validation check using a database engine. In the event of a syntax error, a correction loop is triggered. The system feeds the original query, the resulting error message, and the schema context back to the LLM with instructions to fix the error. This loop repeats until a syntactically correct query is produced or a maximum iteration limit is met, making the system resilient to basic generation mistakes.

\subsection{Semantic validation loop}
A syntactically valid query undergoes semantic validation across four dimensions: Column name consistency (C), Granularity and aggregation (G), Join and relationship consistency (J), and Query logic alignment (Q). An LLM-as-a-judge (prompt in Appendix A) scores each dimension, and a final semantic score $S$ is calculated as a weighted sum of scores as shown in \ref{eq:semantic_formula}.
\begin{equation} \label{eq:semantic_formula}
S = w_c \cdot C + w_g \cdot G + w_j \cdot J + w_q \cdot Q 
\end{equation}
where $w_c, w_g, w_j, w_q$ are predefined weights. If $S$ is below threshold $\theta_s$, the query is sent back to the LLM with diagnostic feedback for iterative correction.

\section{Entailment}
Once a query passes all validation, it is executed in database engine to retrieve the results. In a final verification stage, we check if the result set $R$ logically entails the user's original query $Q$.

\subsection{Entailment scoring}
As shown in \ref{eq:entailment_score} an entailment score $E(Q, R)$ is computed by an LLM-as-a-judge (see Appendix A), assessing both \textbf{Schema consistency} ($E_{\text{schema}}$) and \textbf{Data consistency} ($E_{\text{data}}$).
\begin{equation} \label{eq:entailment_score}
E(Q, R) = \beta \cdot E_{\text{schema}} + (1 - \beta) \cdot E_{\text{data}} \quad \text{where } \quad \beta \in [0,1]
\end{equation}

\subsection{Entailment decision}
If $E(Q, R) \geq \theta_e$, the result is returned to the user, and the KB is updated with the successful query-result pair. If not, the entire pipeline is re-triggered with feedback on the failure reason (e.g., missing columns, incorrect filters). This final loop ensures the system's output truly matches the user's intent.

\section{Mitigating Bias in LLM-based Evaluation}
We acknowledge that using an LLM as a judge can introduce potential biases \cite{zheng2023judging}. To mitigate this, our framework employs a structured, few-shot prompting strategy for all scoring tasks. As shown in Appendix A, our prompts provide detailed instructions, scoring rubrics, and few-shot examples. These examples anchor the LLM's judgment to our specific, task-aligned criteria, increasing the reliability and reproducibility of the evaluation while reducing the impact of the model's inherent biases.

\section{Model Parameter Optimization}
\label{sec:parameters}
We conducted a systematic grid search to determine the optimal values for all parameters: thresholds ($\theta_r, \theta_s, \theta_e$) and weights ($\alpha, \beta, w_c, w_g, w_j, w_q$). Each parameter was varied and the combination that yielded the highest accuracy in our validation set was selected. The final optimized values are presented in Table~\ref{tab:optimized-params}.

\begin{table}[ht]
\caption{Optimized Parameters using grid search.}\label{tab:optimized-params}
\centering
\begin{tabular}{ll}
\toprule
\textbf{Parameter} & \textbf{Value} \\
\midrule
$\theta_r$ & 4 \\
$\theta_s$ & 5 \\
$\theta_e$ & 3 \\
$\alpha$   & 0.6 \\
$\beta$    & 0.2 \\
$w_c$      & 0.1 \\
$w_g$      & 0.3 \\
$w_j$      & 0.4 \\
$w_q$      & 0.2 \\
\bottomrule
\end{tabular}
\end{table}

\section{Evaluation}
\subsection{Experimental setup}

We evaluated our framework using the \textbf{BIRD benchmark} \cite{li2023bird}. Given that accessing large-scale, real-world enterprise datasets for academic experiments is often infeasible due to their proprietary nature, the BIRD benchmark serves as an excellent proxy. It is designed to test Text-to-SQL systems over diverse relational databases that mirror real-world complexity, making it a suitable testbed for our system.

\subsection{Results}
\label{subsec:auto-eval}
We evaluated using Claude, OpenAI, DeepSeek API based, and DeepSeek open-source family models. Table~\ref{tab:eval-results} presents the execution accuracy on the BIRD dataset. Among all models, enabling our reflection framework resulted in substantial accuracy gains over evaluation without reflection framework with overall 6-8\% gain in accuracy. These results confirm that our iterative reasoning and refinement mechanism plays a critical role in enhancing performance.

\begin{table}[ht]
\caption{Accuracy (\%) of various models on the BIRD dataset. Top performers in each category are highlighted.}\label{tab:eval-results}
\centering
\resizebox{\columnwidth}{!}{%
\begin{tabular}{lcccc}
\toprule
\textbf{Model} & \textbf{Simple} & \textbf{Moderate} & \textbf{Challenging} & \textbf{Overall} \\
\midrule
\multicolumn{5}{l}{\textbf{Claude Models}} \\
\textbf{Claude-sonnet-4.5 (with reflection)}  & \textbf{88.97} & \textbf{50.65} & \textbf{32.41} & \textbf{72.03} \\
\texttt{Claude-sonnet-4.5 (w/o reflection)}     & 81.18 & 45.25  & 24.13 & 64.92 \\
\midrule
\multicolumn{5}{l}{\textbf{OpenAI API Models}} \\
\textbf{o1-preview (with reflection)}  & \textbf{75.89} & \textbf{48.49} & \textbf{18.62} & \textbf{62.19} \\
\texttt{o1-preview (w/o reflection)}     & 69.35 & 40.54  & 15.86 & 55.66 \\
\midrule
\multicolumn{5}{l}{\textbf{DeepSeek API Models}} \\
\textbf{deepseek-reasoner (with reflection)}   & \textbf{73.62} & \textbf{46.55} & \textbf{17.24} & \textbf{60.10} \\
\texttt{deepseek-reasoner} (w/o reflection)  & 62.90 & 40.54  & 12.41 & 51.88 \\
\midrule
\multicolumn{5}{l}{\textbf{Open Source Models (Local setup)}} \\
\textbf{DeepSeek-Coder-V2-Lite... (with reflection)}    & \textbf{43.55} & \textbf{21.62} & \textbf{3.44} & \textbf{35.02} \\
\texttt{DeepSeek-Coder-V2-Lite...} (w/o reflection)    & 30.65 & 13.51 & 0.0 & 22.64 \\
\bottomrule
\end{tabular}
}
\end{table}

\subsection{Human evaluation}

To complement our automatic evaluation, we conducted a qualitative human assessment to analyze the correctness and semantic alignment of generated queries, particularly focusing on the impact of the entailment check. This evaluation centered on identifying subtle logical errors that might not be caught by syntax or basic semantic validation alone to validate how framework works. Table ~\ref{tab:refined-sql-example} shows one such example.

\begin{table}[ht]
\caption{Example of SQL refinement driven by entailment feedback.}\label{tab:refined-sql-example}
\centering
\small
\begin{tabular}{ >{\bfseries}m{2.2cm} m{10cm} }
\toprule
\textbf{Stage} & \textbf{Analysis / SQL Snippet} \\
\midrule
User Query & List the lowest three eligible free rates for students in continuation schools. \\
\addlinespace 
Initial SQL & \texttt{SELECT ... (f.Count / f.Enrollment) ... WHERE s.SOCType LIKE '\%Continuation\%'} \\
\addlinespace
Entailment Check Result & \textbf{Score}: 2 / 5 \newline \textbf{Reason:} The result returns None values due to division by zero or nulls. Filters for positive enrollment and float casting are needed. \\
\addlinespace
Refined SQL & \texttt{SELECT ... (CAST(f.Count AS REAL) / f.Enrollment) ... WHERE f.Enrollment > 0 AND s.SOCType LIKE '\%Continuation\%'} \\
\bottomrule
\end{tabular}
\end{table}

\subsection{Comparison with State-of-the-Art}
We benchmarked our framework against leading State-of-the-Art methods on the BIRD development set (Table \ref{tab:sota-comparison}). Reflect-SQL achieves a state-of-the-art execution accuracy of 72.03\%, marking a significant improvement over DIN-SQL \cite{pourreza2023din} (50.72\%), DAIL-SQL (57.41\%) and MAC-SQL (59.59\%). This performance is comparable to the recently proposed CHASE-SQL architecture \cite{pourreza2024chase}, establishing Reflect-SQL as a top-tier solution for complex database querying.

\begin{table}[ht]
\caption{Comparison with State-of-the-Art methods on BIRD dev set (Execution Accuracy \%).}\label{tab:sota-comparison}
\centering
\begin{tabular}{lc}
\toprule
\textbf{Method} & \textbf{Overall Accuracy} \\
BIRD paper \cite{li2023bird} & 46.35 \\
DIN-SQL \cite{pourreza2023din} & 50.72 \\
DAIL-SQL \cite{gao2023text} & 57.41 \\
MAC-SQL \cite{wang2024mac} & 59.59 \\
CHASE-SQL + Claude 3.5 Sonnet \cite{pourreza2024chase} & 69.53 \\
CHASE-SQL + Gemini 1.5 Pro \cite{pourreza2024chase} & \textbf{73.01} \\
\midrule
\textbf{Reflect-SQL + Claude 4.5 Sonnet (Ours)} & \textbf{72.03} \\
\bottomrule
\end{tabular}
\end{table}

\section{Ablation Study}
To understand the contribution of each component, we performed an ablation study using deepseek-chat as the base model. As shown in Table~\ref{tab:ablation-study}, removing any component of the Reflect-SQL framework degrades performance. The self-reflection loop in SQL generation proved most critical, with its removal causing the sharpest drop in accuracy. These findings affirm the effectiveness of our multi-stage, reflective approach.

\begin{table}[ht]
\caption{Ablation results showing the contribution of each module.}\label{tab:ablation-study}
\centering
\footnotesize
\resizebox{\columnwidth}{!}{%
\begin{tabular}{lcccc}
\toprule
\textbf{Variant} & \textbf{Simple} & \textbf{Moderate} & \textbf{Challenging} & \textbf{Overall} \\
\midrule
\textbf{Full Model (Self-Reflect)}       & 68.64 & 43.10 & 14.48 & 55.80 \\
w/o Self-Reflection in SQL Generation    & 62.91 & 37.93 & 12.41 & 50.58 \\
w/o Iterative Retrieval                  & 67.02 & 40.30 & 10.34 & 53.58 \\
w/o Entailment Check                     & 64.10 & 41.59 & 13.10 & 52.47 \\
\bottomrule
\end{tabular}
}
\end{table}

\section{Conclusion and Future Work}
In this work, we presented Reflect-SQL, a self-reflection framework that leverages iterative retrieval, generation, and validation to produce accurate SQL queries. Our approach addresses key challenges in large-schema environments by refining user queries and SQL outputs in a closed feedback loop. By continuously updating its knowledge base, the system adapts over time, enhancing its performance for real-world use cases.
In future work, we plan to:

1. Explore graph-based schema representations to better model complex inter-table relationships and improve the efficiency of our hierarchical retrieval.

2. Develop more advanced entailment scoring methods, using techniques like contrastive learning to enhance the framework's ability to detect subtle misalignments between results and user intent.

\section{Limitations}
Our framework's effectiveness depends on the initial quality of the knowledge base and can face scalability challenges with extremely large schemas. Furthermore, the automatic evaluation metrics for semantic correctness and entailment, while effective, are still an evolving area of research.

\newpage
\appendix
\section{Example Prompts for LLM-as-a-judge Scoring}

Prompt for Semantic Scoring
\begin{footnotesize}
\begin{verbatim}
You are an expert SQL analyst. Score a generated SQL query on four
dimensions of semantic correctness (1-10).
**User Query:** "{user_query}"
**Database Schema:** {schema_json}
**Generated SQL:** "{sql_query}"
**Instructions:**
1.  **Column Name Consistency:** Are all columns valid?
2.  **Granularity and Aggregation:** Are aggregations/filters correct?
3.  **Join and Relationship Consistency:** Are joins logical?
4.  **Query Logic Alignment:** Does the logic match the user's intent?
**Few-shot Examples:**[...]
**Current Task:**
Evaluate the provided query, schema, and SQL. Respond in JSON format:{...}
\end{verbatim}
\end{footnotesize}

\section{Knowledge Base Structure}\label{sec:appendix-kb-example}

Figure~\ref{fig:json-schema-example} illustrates the structured JSON format of the constructed Knowledge Base. This representation captures the semantic descriptions of tables and columns derived from the raw schema and auxiliary evidence, serving as the source for vector embedding generation.
\begin{figure}[ht]
\centering
\begin{minipage}{\linewidth}
\footnotesize
\begin{verbatim}
{
  "financial": {
    "account": {
      "table_description": "it stores information about bank accounts ...",
      "relationships": [{
          "column_name": "district_id", "related_table": "district", 
          "related_column": "district_id"
        }...],
      "column_description": {
        "account_id": "Unique identifier for each bank account...",
        "district_id": "Identifier representing the district...",
        ...}
    }...}
  ...
}
\end{verbatim}
\caption{Example JSON specification of the knowledge base}
\label{fig:json-schema-example}
\end{minipage}
\end{figure}

%
%
%

\end{document}